\documentclass[aps,float,prd,twocolumn,psfig,amsmath]{revtex4-2}
\usepackage[dvipdfmx]{graphicx}
\usepackage{amsmath,amssymb,natbib,siunitx,booktabs,url}
\usepackage{color}
 \usepackage[normalem]{ulem}
\useunder{\uline}{\ul}{}

\newcommand{\lsim}{\lesssim}
\newcommand{\gsim}{\gtrsim}

\newcommand{\beq}{\begin{equation}}
\newcommand{\beqa}{\begin{eqnarray}}
		  \newcommand{\eeq}{\end{equation}}
\newcommand{\eeqa}{\end{eqnarray}}

\begin{document}

\title{Temporal Correlation Statistic for Intrinsic Phase Fluctuation in Double White Dwarf Gravitational-Wave Signals}

\author{Naoki Seto }
\affiliation{Department of Physics, Kyoto University, 
Kyoto 606-8502, Japan
}

\date{\today}

\begin{abstract}
We present a framework to probe intrinsic stochastic fluctuation in the orbital phase evolution of long-lived double white dwarf binaries through gravitational-wave observations with LISA. To capture the essential structure of the fluctuation,
we introduce a minimal quadratic statistic that isolates its temporal correlation.
We derive a simple analytic scaling relation for the signal-to-noise ratio of this correlation statistic, explicitly showing its dependence on the total observation time and the intrinsic phase correlation time.

\end{abstract}

\maketitle

\section{Introduction}

Space-based gravitational-wave (GW) observatories like LISA will resolve a large population of compact binaries in the millihertz band \cite{amaro2023astrophysics,hu2017taiji,luo2016tianqin}.
Among these, double white dwarfs (DWDs) are expected to be particularly abundant, with $\sim10^4$ systems identified over the nominal four-year mission  \cite{Korol2017}.
Their long-lived, nearly monochromatic signals enable continuous tracking of orbital phase evolution over the  observational period $T_{\rm obs}\gsim 4$yr \cite{amaro2023astrophysics}.

For such systems, the GW phase provides a direct and well-defined tracer of the binary's orbital motion \cite{blanchet2014gravitational,poisson2014gravity,maggiore2008gravitational}.
Unlike electromagnetic observables, the generation of GWs is directly tied to the bulk orbital dynamics and is insensitive to source-specific effects such as viscous heating. 

In standard GW analyses of long-lived DWDs, the orbital phase evolution is modeled deterministically using the orbital frequency and its low-order time derivatives  (see e.g., \cite{takahashi2002parameter}).
At sufficiently high phase measurement precision, however, DWDs in the LISA band may exhibit small deviations from  idealized smooth phase evolution  \cite{Gokhale2007,FullerLai2012}.
Finite-size effects can couple the orbital dynamics to additional physical processes, possibly introducing small phase perturbations beyond deterministic secular evolution.
Such departures appear as phase residuals once a deterministic model is applied. Timing observations of the semi-detached DWD HM~Cancri over multi-year baselines \cite{Strohmayer2021,Strohmayer2005} may serve as a benchmark for phase stability in interacting systems. While broadly consistent with smooth secular evolution, they do not exclude stochastic phase fluctuation at the level of $\sim 0.1$ radians. We therefore ask how sensitively such fluctuation can be detected in LISA data.

A closely related framework has long been developed in pulsar timing, where deviations from simple deterministic spin-down models are analyzed through correlated timing residuals \cite{Edwards2006,Coles2011}.
In that context, the pulse arrival times of neutron stars are tracked over decades, and residual correlations are used to identify stochastic or long-timescale components beyond secular evolution \cite{Hobbs2006}.
Such methods have played a central role in nanohertz GW searches  \cite{Hellings1983,vanHaasterenLevin2013}.

The present work adopts a viewpoint similar to pulsar timing, but reformulates it for GW observations of DWDs, where phase information is extracted through template-based matched filtering.
We focus on time-domain correlations as a minimal diagnostic of intrinsic
stochastic phase fluctuation.
Our aim is to establish a transparent baseline for assessing such fluctuation, prior to  full covariance or likelihood treatment.
Within this framework, we derive a simple analytic scaling relation that
clarifies how the sensitivity depends on observation time, total signal-to-noise
ratio, and intrinsic correlation timescale.

For concreteness, we adopt a fiducial example consisting of a nearly monochromatic DWD with a GW frequency of 3 mHz, observed over a total duration $T_{\rm obs}=4\,\mathrm{yr}$ with a matched-filter signal-to-noise ratio $\rho_{\rm tot}=100$  \cite{Korol2017,amaro2023astrophysics}.

The structure of this paper is as follows.
In Sec.~\ref{sec:2}, we define segment-level phase residuals and the linear operations applied to the data, including polynomial phase projection.
In Sec.~\ref{sec:3}, we construct a time-domain correlation statistic and derive its expectation value, variance, and scaling behavior under simple assumptions.
The implications for detectability and sensitivity to intrinsic correlation timescales are discussed in Sec.~\ref{sec:4}, followed by related discussions in Sec.~\ref{sec:5}. Sec. \ref{sec:6} is devoted to a brief summary.   Appendix~A discusses the effects of polynomial phase fitting  in more detail.

\section{Polynomial phase fitting as a linear projection}
\label{sec:2}

In this section, we clarify that polynomial phase fitting in long-lived GW signals is equivalent to a linear projection acting on the phase fluctuation.
This equivalence enables a unified treatment of intrinsic phase fluctuation and detector noise.
\subsection{Setup}

We write the observed data stream as
\begin{equation} \label{dgn}
d(t)=g(t)+n(t),
\end{equation}
where $g(t)$ is the GW signal and $n(t)$ denotes detector noise.

We model the GW signal as
\begin{equation} \label{gt}
g(t)=A(t)\,e^{i[\phi_p(t;\boldsymbol a_t)+s(t)]},
\end{equation}
where $A(t)$ is a slowly varying amplitude, and $s(t)$ denotes a small intrinsic phase fluctuation satisfying $|s(t)|\ll 1$.
The secular phase evolution is described by a low-order polynomial,
\begin{equation}
\phi_p(t;\boldsymbol a)=\sum_{m=0}^{p} a_m t^m ,
\end{equation}
where $\boldsymbol a=(a_0,\ldots,a_p)$ denote the model parameters, whose true values are $\boldsymbol a_t$. 

In this paper, we neglect the deterministic annual modulation
induced by the detector's orbital motion
\cite{amaro2023astrophysics,Cutler:1997ta,takahashi2002parameter}.
{This simplification is made to isolate the analytic scaling of the
intrinsic stochastic phase-correlation statistic. In a realistic analysis,
the Doppler, amplitude, and polarization modulations produced by the
annual motion of LISA should be included in the fitted waveform, or
equivalently treated as additional deterministic degrees of freedom in
the projection. Residual phase components that overlap with these
deterministic response modes can therefore be partially absorbed or
distorted by the fit, especially for timescales close to one year. The
resulting sensitivity should be calibrated with the full time-dependent
LISA response.}

\subsection{Fitting with a simplified template}
We introduce a simplified template waveform,
\begin{equation}
h(t;\boldsymbol a)=A(t)\,e^{i\phi_p(t;\boldsymbol a)} ,
\end{equation}
which retains only the secular polynomial phase evolution.
The template parameters are obtained by matched filtering the data, and the fitted values are denoted by $\boldsymbol a_f$.
For a nearly monochromatic signal, the noise spectral density can be approximated as constant across the signal bandwidth. The noise-weighted inner product then reduces, up to normalization, to a simple time-domain integral, rendering the matched-filter problem equivalent to a least-squares fit in the time domain (see e.g.,  \cite{JaranowskiKrolakSchutz1998}). In the present work, we frequently apply this approximation.

Since the template does not model the intrinsic phase fluctuation $s(t)$ and the data contain detector noise, the fitted parameters $\boldsymbol a_f$ generally deviate from their true values $\boldsymbol a_t$.
We define the parameter estimation error as
\begin{equation}
\delta\boldsymbol a \equiv \boldsymbol a_f-\boldsymbol a_t .
\end{equation}
In the small-fluctuation ($s(t)\ll 1$) and high-signal-to-noise regime ($\rho_{\rm tot}\gg 1$), the mismatch can be linearized, yielding
\begin{equation}
\delta\boldsymbol a = \delta\boldsymbol a_s + \delta\boldsymbol a_n ,
\end{equation}
where the two terms denote contributions from the intrinsic phase fluctuation and detector noise, respectively.

Expanding the true  phase to first order in the parameter estimation error $\delta\boldsymbol a$, we have
\begin{equation} \label{ph1}
\phi_p(t;\boldsymbol a_t)+s(t)
\simeq
\phi_p(t;\boldsymbol a_f)+s'(t)+n_c'(t) ,
\end{equation}
where $s'(t)=s(t)- \phi_p\!\left(t;\delta\boldsymbol a_s\right)$ denotes the intrinsic phase fluctuation remaining after phase fitting, and $n_c'(t)=- \phi_p\!\left(t;\delta\boldsymbol a_n\right)$ is an effective phase contribution induced by detector noise through parameter estimation error \cite{CutlerVallisneri2007}.
Phase fitting removes from both $s(t)$ and $n(t)$ the components that are degenerate with the polynomial phase model.

This absorption can be viewed as a linear projection onto the subspace orthogonal to low-order polynomial phase variations \cite{Rao1973,SeberLee2003,CutlerVallisneri2007}. 
In the continuous-time formulation, the corresponding expressions become algebraically involved. 
For clarity, we make this projection structure explicit in the discretized (segment-space) formulation introduced below.

\subsection{Discretization}

To obtain a discretized time series that probes the post-fit phase residuals appearing in Eq.~(\ref{ph1}), we introduce a segment-level phase estimator.
We divide the data stream into $N$ segments of duration $T_{\rm seg}$ and assign a representative time $t_k$ to each segment $(k=1,\ldots,N)$.

Using the fitted polynomial-phase template $h(t;\boldsymbol a_f)$, we define a segment-level estimator
\begin{equation} \label{xk}
x_k \equiv
\frac{\mathrm{Im}\,(d|h_k(\boldsymbol a_f))}
     {(h_k(\boldsymbol a_f)|h_k(\boldsymbol a_f))},
\end{equation}
where $(\cdot|\cdot)$ denotes the usual noise-weighted inner product and
$h_k(\boldsymbol a_f)$ is the restriction of the template to segment $k$ (see also \cite{JaranowskiKrolakSchutz1998,CutlerVallisneri2007}).
As can be readily verified from a first-order expansion of the complex exponential in Eq.~(\ref{ph1}), this estimator isolates the phase deviations at leading order.
It corresponds to projecting the data onto the quadrature component orthogonal to the fitted template
\footnote{
In a real-valued formulation, this operation is equivalent to taking an inner
product with a template whose phase is shifted by $\pi/2$ relative to the
original template $h$, i.e., projecting onto the quadrature component that is
orthogonal to $h$ in the real signal space.
}.

Under the conditions $s(t)\ll  1$ and $\rho_{\rm tot}\gg 1$, the segment-level estimator can be written as
\begin{equation}\label{xk2}
x_k \simeq s_k' + n_k' .
\end{equation}
Here
\begin{equation}\label{skd}
s_k' \equiv
\frac{(s'h_k(\boldsymbol a_f)|h_k(\boldsymbol a_f))}
     {(h_k(\boldsymbol a_f)|h_k(\boldsymbol a_f))}
\end{equation}
denotes the contribution from the intrinsic phase fluctuation.

In Eq. (\ref{xk2}), 
the noise contribution is given by
\begin{equation}
n_k' = n_k + n_{c,k}' ,
\end{equation}
where the direct detector-noise term and the fitting-induced contribution are defined as
\begin{align}
n_k &\equiv
\frac{\mathrm{Im}\,(n|h_k(\boldsymbol a_f))}
     {(h_k(\boldsymbol a_f)|h_k(\boldsymbol a_f))}, \label{n1} \\
n_{c,k}' &\equiv
\frac{(n_c'h_k(\boldsymbol a_f)|h_k(\boldsymbol a_f))}
     {(h_k(\boldsymbol a_f)|h_k(\boldsymbol a_f))}. \label{n2}
\end{align}
By construction, $s_k'$ and $n_k'$ represent the intrinsic and
detector-noise contributions to the phase estimate $x_k$ in segment $k$.
In Eq.~(\ref{xk2}), they are not separately observable; rather,
they constitute a decomposition of the single measured quantity
$x_k$ into physically distinct contributions.
This decomposition  clarifies the two origins of the segment-level phase deviations and provides a convenient basis for the statistical analysis developed below.

We also define
\begin{equation}
s_k \equiv
\frac{(sh_k(\boldsymbol a_f)|h_k(\boldsymbol a_f))}
     {(h_k(\boldsymbol a_f)|h_k(\boldsymbol a_f))}  \label{s1}
\end{equation}
as the segment-level contribution associated with the original fluctuation $s(t)$.

\subsection{Evaluation of the intrinsic contribution}
Next
we  explicitly evaluate the intrinsic contribution $s_k'$ introduced above.
From Eqs.~(\ref{ph1}) and (\ref{skd}), this quantity provides a discretized description of the intrinsic phase fluctuation after polynomial phase fitting.

When the segment duration satisfies $T_{\rm seg}\ll 1\,\mathrm{yr}$, variations in the signal amplitude $A(t)$ on annual timescales can be neglected.
In this regime, the intrinsic contribution reduces to a simple time average of the post-fit residual phase over each segment 
\begin{equation} \label{sk'}
s_k' \simeq \frac{1}{T_{\rm seg}}
\int_{\text{segment }k} s'(t)\,dt .
\end{equation}
This relation provides the leading-order mapping from the continuous-time intrinsic fluctuation $s'(t)$ to the discretized time series used in the subsequent analysis.
An analogous expression holds for $s_k$, defined in terms of the original fluctuation $s(t)$.

For compact notation, we collect the segment-level intrinsic contributions into vector form,
\begin{equation}
\boldsymbol{s}' \equiv (s_1', s_2', \ldots)^{\mathsf T},
\qquad
\boldsymbol{s}\equiv (s_1, s_2, \ldots)^{\mathsf T}.
\end{equation}

The effect of polynomial phase fitting on the intrinsic fluctuation can be expressed as a linear operation in the segment space (see e.g., \cite{Coles2011,vanHaasterenLevin2013}).
Let $X$ denote the design matrix whose columns are given by the polynomial basis functions evaluated at the representative segment times $t_k$,
\begin{equation}
X_{k m} = t_k^{m}, \qquad m=0,1,\ldots,p .
\end{equation}
In this representation, the intrinsic fluctuation sequence after phase fitting is given by
\begin{equation}\label{ps}
\boldsymbol{s}' = P\,\boldsymbol{s},
\end{equation}
with the projection operator \cite{Coles2011,vanHaasterenLevin2013}
\begin{equation}\label{P}
P = I - X (X^{\mathsf T} X)^{-1} X^{\mathsf T}.
\end{equation}
This projection structure is equivalent to the standard linear estimator associated with polynomial parameter fitting and is closely related to the Fisher-matrix formulation in the high-SNR limit (see e.g., \cite{CutlerVallisneri2007}).
It removes components of $\boldsymbol{s}$ that are degenerate with the polynomial phase model, retaining only fluctuation beyond smooth secular trends.

\subsection{Detector noise contribution}

We now consider the contribution of detector noise to the segment-level phase estimator, $n_k' = n_k + n_{c,k}'$ [cf.\ Eqs.~(\ref{n1}) and (\ref{n2})].

Analogously to Eq.~(\ref{sk'}), the fitting-induced noise contribution can be approximated as
\begin{equation}
n_{c,k}' \simeq \frac{1}{T_{\rm seg}}
\int_{\text{segment }k} n_c'(t)\,dt .
\end{equation}
It is convenient to collect the noise-related quantities into vector form,
\begin{equation}
\boldsymbol n' \equiv (n_1', n_2', \ldots)^{\mathsf T},
\qquad
\boldsymbol n \equiv (n_1, n_2, \ldots)^{\mathsf T}.
\end{equation}

In the discretized representation, we find that the effect of phase fitting on detector noise can again be written as a linear operation,
\begin{equation} \label{pn}
\boldsymbol n' = P\,\boldsymbol n ,
\end{equation}
where $\boldsymbol n'$ denotes the noise contribution after polynomial phase fitting, and $P$ is the same projection operator defined previously.

Formally, Eqs.~(\ref{ps}) and (\ref{pn}) take a compact form as if the data were first discretized into segment-level quantities and subsequently projected by the operator $P$.
It should be stressed, however, that this interpretation does not reflect the actual order of operations in the derivation.
In practice, we evaluate the effect of parameter-estimation errors within the continuous-time matched-filtering framework, and construct the segment-level estimators only afterward, under standard linearization and discretization approximations.
Projection representations (\ref{ps}) and (\ref{pn}) therefore provide a convenient linear description of the final result, rather than a shortcut in the derivation.
At linear order, this rearrangement is justified by the fact that discretization through inner products and the projection induced by phase fitting are both linear operations and hence commute, leading to equivalent results irrespective of the order in which they are applied.

\section{Statistical structure of projected phase residuals}   
\label{sec:3}
In this section, we characterize the correlation structure of the segment-level projected phase residuals, for both detector noise $n_k'$ and intrinsic phase fluctuation $s_k'$.

This comparison clarifies their statistical distinction, which underlies the non-diagonal correlation statistic introduced in the next section.

\subsection{Detector-noise correlations before and after projection}
\label{sec:3.1}
We first evaluate the covariance $\langle n_k' n_{k+m}' \rangle$ of the projected detector noise.

We denote by $n_k$ the detector-noise contribution prior to polynomial
projection, which satisfies
\begin{equation}
\langle n_k \rangle = 0 .
\end{equation}
A finite segment duration induces small inter-segment
correlations.
For stationary noise, these scale as
\begin{equation}
\frac{\langle n_k n_{k+m} \rangle}{\langle n_k^2 \rangle}
= \mathcal O\!\left(\frac{1}{f\,m\,T_{\rm seg}}\right),
\qquad (m\ge 1). 
\end{equation}
These correlations are negligible for $fT_{\rm seg}\gg 1$, and we adopt a white-noise approximation at the segment level
\begin{equation}
\langle n_k n_{k+m} \rangle \simeq \sigma_n^2\,\delta_{m0}. 
\label{eq:nk_white}
\end{equation}

From Eq. (\ref{n1}), 
the variance $\sigma_n^2$ is set by the signal-to-noise ratio $\rho_{\rm seg}$  accumulated within
a single segment.
Using a phase estimator constructed from a single quadrature of the
matched-filter output, we estimate
\begin{equation}
\sigma_n^2 \simeq \frac{1}{2\rho_{\rm seg}^2},
\qquad
\rho_{\rm seg}
= \rho_{\rm tot}\sqrt{\frac{T_{\rm seg}}{T_{\rm obs}}}.
\end{equation}
For the fiducial model adopted throughout this work, we take
$\rho_{\rm tot}=100$ for $T_{\rm obs}=4\,\mathrm{yr}$ and a default segment
duration $T_{\rm seg}=20\,\mathrm{days}$, corresponding to
$\rho_{\rm seg}\simeq 10$ and $N\simeq 70$.
These values are introduced only to characterize the scaling of the noise
amplitude and do not enter the formal derivations below.

We now consider the effect of polynomial phase fitting on the detector noise $n_k$.
As shown in Sec.~\ref{sec:2}, the associated projection  acts linearly on the
segment-level quantities,
\begin{equation}
\boldsymbol n' = P\,\boldsymbol n ,
\end{equation}
where $P$ denotes the projection operator onto the subspace orthogonal to the
polynomial phase model.
The operator $P$ is symmetric and idempotent ($P^{\mathsf T}=P$ and $P^2=P$).
Assuming the pre-projection covariance~(\ref{eq:nk_white}), the covariance of the
projected detector noise is
\begin{equation}\label{nd2}
\langle n_k' n_{k+m}' \rangle
= \sigma_n^2\, P_{k,k+m}.
\end{equation}

For a polynomial model of fixed order applied to $N$ segments, the diagonal
elements satisfy $P_{kk}=1-\mathcal O(p/N)$, while the off-diagonal elements are
suppressed as
\begin{equation}
P_{k,k+m}=\mathcal O(p/N),
\qquad (m\neq 0).
\end{equation}
Details of this scaling are discussed in Appendix~A.

We conclude that detector noise induces only weak non-diagonal correlations in
the projected segment-level residuals, defining a well-controlled null
hypothesis for the non-diagonal correlation analysis developed in the following
section.

\subsection{Intrinsic phase fluctuation and its correlation}
\label{sec:3.2}
We now turn to the  covariance $\langle s_k' s_{k+m}' \rangle$ of the projected intrinsic fluctuation.

Before the projection, 
we assume that the intrinsic phase fluctuation has vanishing mean,
\begin{equation}
\langle s_k \rangle = 0 ,
\end{equation}
and are statistically stationary  across segments. We also define
\begin{equation}
\langle s_k^2 \rangle = \sigma_s^2 .
\end{equation}
Their second-order statistics are  generally characterized by the lag correlation function
\begin{equation}
C_m \equiv \langle s_k\, s_{k+m} \rangle ,
\end{equation}
which depends only on the separation $m$ between segments.

To characterize the temporal extent of the intrinsic correlations, we introduce
a correlation time scale $\tau_{\rm int}$.
In terms of the segment lag $m$, this corresponds to an effective correlation
length
\begin{equation}
m_{\rm int} \sim \frac{\tau_{\rm int}}{T_{\rm seg}} .
\end{equation}
As a working approximation, we expect
\begin{equation}
C_m \sim C_0=\sigma_s^2
\quad \text{for} \quad
m \lesssim m_{\rm int},
\end{equation}
while $C_m$ decays for $m \gg m_{\rm int}$.
No specific functional form for this decay is assumed.
This provides a model-independent characterization of the range over which
intrinsic phase fluctuation exhibits non-diagonal correlations.

We next consider the effect of the polynomial phase fitting on $s_k$. As presented in Eq. (\ref{ps}), we have
\begin{equation}
\boldsymbol s' = P\,\boldsymbol s .
\end{equation}

The correlation function of the projected intrinsic fluctuation is therefore
given by
\begin{equation}
\langle s_k' s_{k+m}' \rangle
= \sum_{i,j} P_{k i}\, P_{k+m,j}\, \langle s_i s_j \rangle ,
\end{equation}
which may be written schematically, using stationarity, as
\begin{equation}
\langle s_k' s_{k+m}' \rangle
= (P C P)_{k,k+m},
\end{equation}
where $C$ denotes the covariance matrix with elements
$C_{ij} = C_{|i-j|}$.

The projection removes only those components of the intrinsic fluctuation that
are degenerate with the low-order polynomial phase evolution over the full
observation time $T_{\rm obs}$.
It therefore modifies the correlation structure on scales
$\lesssim \tau_{\rm int}$ only weakly in the regime
$\tau_{\rm int}\ll T_{\rm obs}$,
as illustrated in Appendix~A for a representative
Ornstein--Uhlenbeck process \cite{Gardiner2009}.
We therefore approximate
\begin{equation}\label{cs}
\langle s_k' s_{k+m}' \rangle 
\sim C_m \sim C_0=\sigma_s^2 ,
\end{equation}
for lags $m \lesssim m_{\rm int}$.
This is in contrast to the nearly diagonal structure of the projected detector noise
[cf.\ Eq.~(\ref{nd2})].

Even if intrinsic phase fluctuation is subdominant to detector noise at the
level of individual segments, $\sigma_s \ll \sigma_n$, their correlated nature
allows non-diagonal statistics to accumulate contributions over
$\mathcal O(m_{\rm int})$ lags.
This motivates the use of the non-diagonal correlation statistic introduced in the
next section.

\section{Non-diagonal correlation statistic}
\label{sec:4}
In this section, we introduce a simple quadratic statistic $Q$
designed to probe intrinsic stochastic phase fluctuation through its temporal correlation.

\subsection{Definition of the statistic}
\label{sec:nd_setup}

We build on the segmented phase estimator expressed as 
\begin{equation}
x_k \simeq  n_k' + s_k' ,
\end{equation}
where $n_k'$ and $s_k'$ denote the detector-noise and intrinsic stochastic
contributions after polynomial phase projection.

Because $\langle x_k\rangle=0$, linear statistics of $x_k$ cannot probe intrinsic phase fluctuation.
The intrinsic contributions $s_k'$ instead manifest themselves in inter-segment two-point functions.
We therefore construct quadratic combinations of different time segments to accumulate short-lag correlations, while suppressing the nearly diagonal detector-noise component $n_k'$.

To make the scaling structure explicit, we adopt a simple top-hat weighting
scheme and define
\begin{equation}
Q
=
\sum_{m=1}^{M}
\sum_{k=1}^{N-m}
x_k\,x_{k+m},
\label{eq:Q_tophat}
\end{equation}
where $N$ is the total number of segments and $M$ is the maximum lag included.
The corresponding accumulation timescale is
\begin{equation}
\tau_2 \equiv M\,T_{\rm seg}.
\end{equation}

\subsection{Noise baseline}
\label{sec:nd_noise}

Under the noise-only hypothesis ($x_k=n_k'$), the expectation value becomes
\begin{equation}
\langle Q \rangle_n
=
\sum_{m=1}^{M}
\sum_{k=1}^{N-m}
\langle n_k' n_{k+m}' \rangle .
\end{equation}

As discussed in Sec.~\ref{sec:3.1}, the off-diagonal correlations of the projected
detector noise are suppressed by $\mathcal{O}(1/N)$,
\begin{equation}
\langle n_k' n_{k+m}' \rangle
=
\sigma_n^2 P_{k,k+m}
=
\mathcal{O}\!\left(\frac{p\sigma_n^2 }{N}\right).
\end{equation}

Summing over $k$ yields
\begin{equation}
\langle Q \rangle_n
\sim
pM \sigma_n^2 ,
\end{equation}
which is much smaller than the statistical fluctuation $\sqrt{\mathrm{Var}_n(Q)}$ (see Eq. (\ref{eq:VarQ_tophat}) below) for $pM \ll N$. Below we approximate
\begin{equation}
\langle Q \rangle_n \simeq 0
\end{equation}
at leading order.

The variance under the noise-only hypothesis is
\begin{equation}
\mathrm{Var}_n(Q)
=
\sum_{m=1}^{M}
\sum_{k=1}^{N-m}
\sum_{m'=1}^{M}
\sum_{k'=1}^{N-m'}
\langle n_k' n_{k+m}' n_{k'}' n_{k'+m'}' \rangle .
\end{equation}

Assuming effective independence between distinct segments at leading order,
the four-point function reduces to products of two-point functions,
yielding
\begin{equation}
\mathrm{Var}_n(Q)
\sim
N M\,\sigma_n^4 ,
\label{eq:VarQ_tophat}
\end{equation}
up to factors of order unity.

Projection-induced correlations modify this result only at relative order
$\mathcal{O}(pM/N)$ and are neglected at leading order.

\subsection{Expectation value from intrinsic correlations}
\label{sec:nd_expectation}

Next, we consider the contribution of the intrinsic component $s_k'$, based on Eq. (\ref{cs}).
For the maximum lag $M\lsim m_{\rm int}$, 
the expectation value scales as
\begin{equation}
\langle Q \rangle
=
\sum_{m=1}^{M}
\sum_{k=1}^{N-m}
C_m
\sim
N M\,\sigma_s^2 .
\label{eq:Q_expectation}
\end{equation}

Even for $\sigma_s \ll \sigma_n$ at the segment level,
the coherent accumulation over $NM$ pairs can enhance the signal.

\subsection{Signal-to-noise ratio and scaling}
\label{sec:nd_sn}

We now combine the results of the previous two subsections in the regime
$\tau_2 \lesssim \tau_{\rm int}$ (equivalently $M \lesssim m_{\rm int}$).
We define the signal-to-noise ratio of the statistic $Q$ as
\begin{equation}
\mathrm{SN}_Q
\equiv
\frac{\langle Q \rangle}{\sqrt{\mathrm{Var}_n(Q)}} .
\end{equation}
With the leading-order scalings
$\langle Q \rangle \sim N M\,\sigma_s^2$
and
$\mathrm{Var}_n(Q) \sim N M\,\sigma_n^4$,
we obtain
\begin{equation}
\mathrm{SN}_Q
\sim
\sqrt{N M}\,
\frac{\sigma_s^2}{\sigma_n^2},
\label{eq:SN_basic}
\end{equation}
up to factors of order unity.

To make the scaling in Eq.~(\ref{eq:SN_basic}) explicit in terms of
observational parameters, we write
\begin{equation}
N = \frac{T_{\rm obs}}{T_{\rm seg}},
\qquad
M = \frac{\tau_2}{T_{\rm seg}},
\end{equation}
where $T_{\rm obs}$ is the total observation time,
$T_{\rm seg}$ the segment duration, and
$\tau_2$ the maximum accumulation timescale of the statistic.

To proceed, we apply the following relations
\begin{equation}
\sigma_n^2 \simeq \frac{1}{2\rho_{\rm seg}^2},
\qquad
\rho_{\rm seg}^2 \simeq \Gamma\,T_{\rm seg},
\end{equation}
where $\Gamma$ denotes the accumulation rate of signal-to-noise of  the original DWD signal presented in Eq.  (\ref{gt}).
The total coherent signal-to-noise ratio   satisfies
\begin{equation}
\rho_{\rm tot}^2 \simeq \Gamma T_{\rm obs}.
\end{equation}

Substituting these relations into Eq.~(\ref{eq:SN_basic}), we obtain
\begin{equation}
\mathrm{SN}_Q
\sim
\sqrt{T_{\rm obs}\,\tau_2}\,
\Gamma\,\sigma_s^2
\sim
\rho_{\rm tot}^2\,\sigma_s^2\,
\sqrt{\frac{\tau_2}{T_{\rm obs}}},
\label{eq:SN_Tobs}
\end{equation}
up to numerical factors of order unity.
Strictly speaking, intrinsic phase fluctuation $s(t)$ reduces the coherent
matched-filter signal-to-noise ratio $\rho_{\rm tot}$ at order $\sigma_s^2$.
In the small-fluctuation regime considered here,
this correction is subleading and can be neglected
when deriving the  scaling relations below.

An important feature of Eq.~(\ref{eq:SN_Tobs}) is the absence of explicit
dependence on $T_{\rm seg}$.
Decreasing $T_{\rm seg}$ increases the number $MN$  of contributing segment pairs,
but simultaneously enlarges the variance $\sigma_n$ of each segment-level estimator,
leading to a cancellation at leading order.

Accordingly, the method does not rely on individually significant
segment-level detections.
What is required is a sufficiently large global matched-filter
signal-to-noise ratio $\rho_{\rm tot}$ accumulated over the full observation time $T_{\rm obs}$.

{Realistic LISA data streams will not be perfectly continuous, and data gaps modify the above idealized scaling through the observing window. The definition of the segment-level estimator in Eq.~(\ref{xk}) can be generalized by evaluating the inner products with the actual observing window within each segment. In this case, gaps reduce the effective signal-to-noise ratio of the affected segments and lead to nonuniform segment weights. At the level of the quadratic statistic, the uniform sum in Eq.~(\ref{eq:Q_tophat}) should be replaced by a weighted sum over available segment pairs. Short or randomly distributed gaps mainly reduce the effective number of contributing pairs, whereas long or regularly spaced gaps can introduce a nontrivial lag dependence in the pair count. In practical applications, the gap-free factor $NM$ should therefore be replaced by the effective number of weighted pairs, and the variance of $Q$ should be calibrated using the actual observing window.}

\subsection{Dependence on the intrinsic correlation time}
\label{sec:nd_tauint}

The lag dependence of $Q$ directly probes the intrinsic correlation time $\tau_{\rm int}$.
For $\tau_2\lsim \tau_{\rm int}$, intrinsic contributions add coherently, yielding $\langle Q\rangle \propto \tau_2$.
Once $\tau_2 \gtrsim \tau_{\rm int}$, correlations no longer persist over the
full range of lags and the growth of $\langle Q\rangle$ saturates.
The onset of this saturation therefore provides an empirical estimate of
$\tau_{\rm int}$.

While the signal saturates beyond $\tau_{\rm int}$, the noise variance continues
to increase with the number of included pairs,
$\mathrm{Var}_n(Q)\propto N M$.
As a result, the signal-to-noise ratio $SN_Q$ is maximized around
$\tau_2 \simeq \tau_{\rm int}$.
Using Eq.~(\ref{eq:SN_Tobs}) with $\tau_2=\tau_{\rm int}$, we obtain
\begin{equation}
\mathrm{SN}_{Q,\max}
\sim
\rho_{\rm tot}^2\,\sigma_s^2\,
\sqrt{\frac{\tau_{\rm int}}{T_{\rm obs}}}.
\label{eq:SN_rhotot}
\end{equation}

The amplitude $\sigma_s$ can be estimated from the measured magnitude of
$\langle Q\rangle$, given the known values of $N$ and $M$.

For the two parameters $\tau_{\rm int}$ and $\sigma_s$, a rough estimate of the attainable precision may be obtained by comparing the expected signal amplitude with the statistical fluctuation of $Q$.
For an order-of-magnitude estimation, this gives
\begin{equation}
\frac{\delta \tau_{\rm int}}{\tau_{\rm int}}
\sim
\frac{\delta \sigma_s^2}{\sigma_s^2}
\sim
\mathrm{SN}_{Q,\max}^{-1}.
\end{equation}

For illustration, let us consider the fiducial system with
$T_{\rm obs}=4\,\mathrm{yr}$ and $\rho_{\rm tot}=100$.
The above scaling then gives
\begin{equation}
\mathrm{SN}_{Q,\max}
\sim
35\,
\left(\frac{\rho_{\rm tot}}{100}\right)^2
\left(\frac{\sigma_s}{0.1}\right)^2
\left(\frac{\tau_{\rm int}}{0.5\,\mathrm{yr}}\right)^{1/2}
\left(\frac{T_{\rm obs}}{4\,\mathrm{yr}}\right)^{-1/2},
\end{equation}
indicating that even modest intrinsic phase fluctuation can produce a statistically significant non-diagonal signal when correlations persist over long timescales.
{The numerical choice $\tau_{\rm int}=0.5\,\mathrm{yr}$ in this
estimate is intended as an illustrative correlation timescale in the
simplified response model. It should not be interpreted as implying that
annual-response effects are negligible near this timescale, especially
when the observing baseline spans only a few annual cycles.}

\section{Discussion}
\label{sec:5}
We now discuss the relation of the proposed statistic to existing methods and its possible extensions.

\subsection{Relation to pulsar timing analyses}
The analysis presented here is conceptually related to studies of intrinsic
noise processes in pulsar timing, where a phase-like observable is tracked over
long timescales and deviations from a low-order polynomial model are analyzed
through timing residuals (e.g., \cite{Coles2011,vanHaasterenLevin2013,Lentati2013}).
In both contexts, the essential task is to identify temporally correlated
fluctuation in phase residuals obtained after fitting a smooth deterministic
phase evolution.

\subsection{Relation to stochastic background analyses}

The non-diagonal correlation statistic introduced here is structurally
analogous to the correlation analysis for  stochastic
GW background searches
\cite{AllenRomano1999,RomanoCornish2017}.
In both cases, correlated components are extracted through quadratic
combinations of data characterized by a non-trivial two-point function.
A key distinction is that the present analysis forms correlations between
different times within a single data stream, rather than between distinct
detectors at the same time.

From this perspective, combinations of noise independent
detector channels, as extensively studied in stochastic background
analyses, are naturally compatible with the present framework.
Examples include combinations of the multiple TDI variables within LISA,
as well as cross-mission networks such as LISA-Taiji.

{The same viewpoint also suggests a population-level implication.
If intrinsic phase fluctuations are present in many DWDs, subtraction
with smooth deterministic templates would leave residuals that are not
solely due to ordinary parameter-estimation errors.  Even after the best-fit
deterministic parameters are removed, the unmodeled stochastic component
can remain as a source-origin residual with temporal correlations inherited
from the underlying phase fluctuation.  Such residuals could contribute to
the post-subtraction foreground, although quantifying this effect would
require a dedicated population-level simulation.}

\subsection{Sensitivity to general phase modulations}
Although the present analysis has focused on internal processes for
isolated DWDs, the same framework can serve as a screening tool for more
general phase modulations. Weak perturbations, such as those induced by
a third body (e.g., \cite{Seto2008,Tamanini2020}) or other external
influences, may generate small but temporally correlated deviations from
a smooth phase evolution.

{
For a well-specified deterministic perturbation, a dedicated coherent
template fit would ultimately provide the most powerful detection and
parameter-estimation method. The role of the present statistic is
complementary: it tests, in a model-independent way, whether the
residuals left after the baseline deterministic phase fit contain
non-diagonal temporal correlations that justify a more detailed model
extension. In this sense, a significant detection of correlated phase
residuals would identify systems that warrant subsequent residual-space
analyses based on physical templates or covariance models.
}

\subsection{Synergies with electromagnetic observations}

The analysis presented here relies exclusively on GW data to probe stochastic features in orbital phase evolution. Electromagnetic observations provide complementary information, and joint interpretation may offer additional insight into the underlying astrophysical processes. {They can also help control deterministic detector-response effects. In particular, the dominant annual Doppler phase modulation is largely determined by the sky position of the binary and the known orbital motion of LISA. For optically identified binaries with accurately measured sky positions, this component can therefore be modeled with high precision, reducing its degeneracy with source-side phase variations. In favorable systems, electromagnetic timing or photometric monitoring may also provide an independent reference for interpreting correlated GW phase residuals.}

\subsection{Applicability to other long-lived compact binaries}

The framework is not limited to DWDs.
Any long-lived compact binary with a coherently trackable
GW phase may be analyzed in a similar manner.
Examples include early inspiral stellar-mass black holes
(e.g., \cite{Seto:2022xmh}) and neutron-star--white-dwarf systems \cite{amaro2023astrophysics}.
A detailed assessment of correlated phase fluctuation in these systems
is left for future work.

\subsection{Modeling Scope and Outlook}

The present analysis isolates the leading non-diagonal correlation structure
induced by stochastic phase fluctuation.
More refined treatments could incorporate explicit covariance modeling
or embed the same structure within a Bayesian inference framework,
as commonly done in pulsar timing analyses and stochastic background searches.
{A full demonstration with simulated LISA data would be a natural next step. Such an analysis would have to include the time-dependent LISA response, parameter covariance, data gaps, and residual foreground or source-subtraction effects. We leave this more complete implementation for future work, while the present paper provides the analytic baseline for the corresponding residual-correlation statistic.}

\section{Summary}
\label{sec:6}

We present a minimal diagnostic framework for intrinsic stochastic
phase fluctuation in GW signals from DWDs,
constructed from time correlations of phase residuals.

Dividing the data into short segments, we form a quadratic
inter-segment correlation statistic.
Analytic expressions for its expectation value and variance show that
the resulting signal-to-noise ratio is independent of the chosen segment duration
and scales with the total observation time and the
effective correlation time.

These results clarify how weak but temporally correlated phase
fluctuation can accumulate coherently over long observations.
They provide a transparent baseline for assessing the detectability
of such effects in nearly monochromatic GW signals from long-lived compact binaries.

\appendix

\section{Effects of polynomial phase fitting}
\label{sec:a1}
This appendix summarizes the statistical effect of polynomial phase
fitting on segment-level detector noise $n_k$ and intrinsic phase
fluctuation $s_k$.

Although both are acted upon by the same linear projection operator $P$,
their statistical responses differ.
For initially uncorrelated detector noise, the induced correlations
are weak and the variance reduction is small.
For intrinsically correlated phase fluctuation, the impact of the
projection depends on the ratio $\tau_{\rm int}/T_{\rm obs}$.

\subsection{Detector noise}
\label{app:projection_noise}

Before phase fitting, the segment-level noise sequence $n_k$
is assumed to be stationary, zero-mean, and uncorrelated,
\begin{equation}
\langle n_k \rangle = 0, \qquad
\langle n_k n_l \rangle
= \sigma_n^2\,\delta_{kl},
\end{equation}
where $\sigma_n^2$ denotes the variance per segment.

Polynomial phase fitting acts on the discretized noise through the
linear projection operator $P$ introduced in the main text,
\begin{equation}
\boldsymbol n' = P\,\boldsymbol n,
\end{equation}
with $P^{\mathsf T}=P=P^2$.
The covariance matrix of the projected noise is therefore
\begin{equation}
\langle \boldsymbol n' \boldsymbol n'^{\mathsf T} \rangle
= \sigma_n^2\, P .
\label{eq:noise_cov_proj}
\end{equation}

Polynomial fitting induces weak inter-segment correlations through
the off-diagonal elements of $P$.
The mean variance reduction is quantified by
\begin{equation}
\frac{\langle n_k' n_k' \rangle}{\langle n_k n_k \rangle}
= \frac{1}{N}\,\mathrm{Tr}(P),
\end{equation}
where $N$ is the number of segments.

For a degree-$p$ polynomial fit,
\begin{equation}
\mathrm{Tr}(P)=N-(p+1),
\end{equation}
implying
\begin{equation}
R_n = 1-\frac{p+1}{N}.
\end{equation}

For $k\neq l$, the induced correlations satisfy
\begin{equation}
\langle n_k' n_l' \rangle
= \mathcal{O}\!\left(\frac{p+1}{N}\right)
\langle n_k'^2 \rangle ,
\end{equation}
reflecting the fact that $P_{kl}=\mathcal{O}((p+1)/N)$ while
$P_{kk}=1-\mathcal{O}((p+1)/N)$.

Polynomial fitting induces $\mathcal{O}(1/N)$ long-range correlations
in the discretized noise.

\subsection{Intrinsic phase fluctuation}
\label{app:projection_signal}

We next consider intrinsic stochastic phase fluctuation.
The fitting again acts through the same projection operator,
\begin{equation}
\boldsymbol{s}' = P\,\boldsymbol{s},
\end{equation}
but the resulting impact depends on the temporal covariance
of the intrinsic process.
We assume $\langle s_k\rangle=0$ and denote
\begin{equation}
\langle s_k s_{k+m}\rangle = C_{|m|}.
\end{equation}

After projection, the covariance becomes
\begin{equation}
\langle \boldsymbol{s}'\boldsymbol{s}'^{\mathsf T} \rangle
= P C P ,
\end{equation}
where $C_{ij}=C_{|i-j|}$.
Unlike the detector-noise case, the suppression therefore depends
explicitly on the structure of $C$.

To quantify the suppression, we define
\begin{equation}
R_s \equiv
\frac{\langle s_k'^2\rangle}{\langle s_k^2\rangle}
=
\frac{\mathrm{Tr}(P C P)}{\mathrm{Tr}(C)} ,
\label{eq:Rs_def}
\end{equation}
which depends on the temporal structure of $C$.

As an illustration, we adopt a discrete Ornstein--Uhlenbeck model with
\begin{equation}
C_{ij} =
\sigma_s^2
\exp\!\left(-\frac{|i-j|\,T_{\rm seg}}{\tau_{\rm int}}\right),
\end{equation}
where $\tau_{\rm int}$ is the intrinsic correlation time.
This example is used only for illustration.

\begin{figure}
  \centering
  \includegraphics[width=0.95\linewidth]{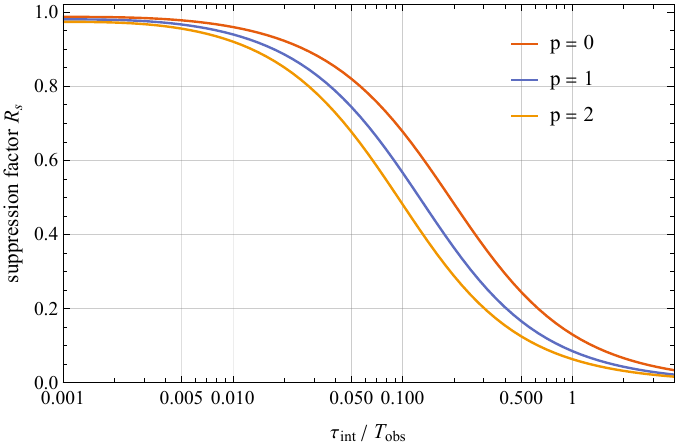}
 \caption{
Suppression factor $R_s$ as a function of
$\tau_{\rm int}/T_{\rm obs}$.
Short-memory fluctuation
($\tau_{\rm int}\ll T_{\rm obs}$) is weakly affected,
whereas long-memory fluctuation is increasingly suppressed.
}
  \label{fig:Rp_suppression}
\end{figure}

For this model, $R_s$ can be evaluated numerically as a function of
$\tau_{\rm int}/T_{\rm obs}$ and $p$.
Figure~\ref{fig:Rp_suppression} shows that
$R_s\simeq 1$ for $\tau_{\rm int}\ll T_{\rm obs}$,
whereas long-memory fluctuation is increasingly suppressed.
Since polynomial fitting is performed over the full observation span,
this qualitative behavior is physically expected and should hold
quite generally, without relying on the specific form of the
intrinsic correlation function.

\bibliography{ref}

\end{document}